\documentclass{iaus}
\usepackage{graphicx}

\title[Star Formation Histories] %% give here short title %%
{Star formation histories of resolved galaxies}

\author[M. Tosi]   %% give here short author list %%
{Monica Tosi
}

\affiliation{INAF - Osservatorio Astronomico di Bologna \\ Via Ranzani 1,
I-40127, Bologna, Italy \\ email: {\tt monica.tosi@oabo.inaf.it} 
}

\pubyear{2008}
\volume{258}  %% insert here IAU Symposium No.
\pagerange{}
\jname{The Ages of Stars}
\editors{E.E. Mamajek, D.R. Soderblom \& R.F.G. Wyse, eds.}
\begin{document}

\maketitle

\begin{abstract}
The colour-magnitude diagrams of resolved stellar populations are the best tool to
study the star formation histories of the host galactic regions. In this review
the method to derive star formation histories by means of synthetic 
colour-magnitude diagrams is briefly outlined, and the results of its application
to resolved galaxies of various morphological types
are summarized. It is shown that all the galaxies studied so far were
 already forming stars at the lookback time reached by the observational
data, independently of morphological type and metallicity. Early-type
galaxies  have formed stars predominantly, but in several cases not 
exclusively, at the earliest epochs. All the other galaxies appear to have
experienced rather continuous star formation activities throughout their
lifetimes, although with significant rate variations and, sometimes,
short quiescent phases.
\keywords{galaxies: evolution, galaxies: stellar content
}
\end{abstract}

\firstsection % if your document starts with a section,
\section{Introduction}
Two complementary approaches are necessary to understand galaxy evolution: 
on the one hand, we need to develop theoretical models for galaxy formation, 
chemical and dynamical evolution, and on the other hand, we
need to collect as many and as accurate as possible observational data to
constrain such models. In particular, we need to know the masses, chemical
abundances and kinematics of the various galactic components, namely gas, stars
and dark matter; we need to know the star formation history (SFH), the initial
mass function (IMF), etc. In this review our current 
knowledge of the SFHs, as derived from the colour-magnitude diagrams (CMDs) 
of their resolved stellar populations, is summarized.

Resolved stellar populations are the best tracers of the SFH of a galactic
region, and their CMD the best tool to exploit the tracers. This is due 
to the well known circumstance that the location of any
individual star in a CMD is uniquely related to its
mass, age and chemical composition. From the CMD we can thus
disentangle directly these evolution parameters.   In the
case of simple stellar populations, i.e. coeval stars with the same chemical
composition, isochrone fitting is the most frequently used method to infer the
system age. In the case of galaxies, with rather complicated mixtures of
different stellar generations, the age determination is less straightforward,
but their CMDs remain the best means to derive the SFH.  

\section{CMD synthesis and Star Formation Histories}

\begin{figure}
\centering
\includegraphics[width=13cm]{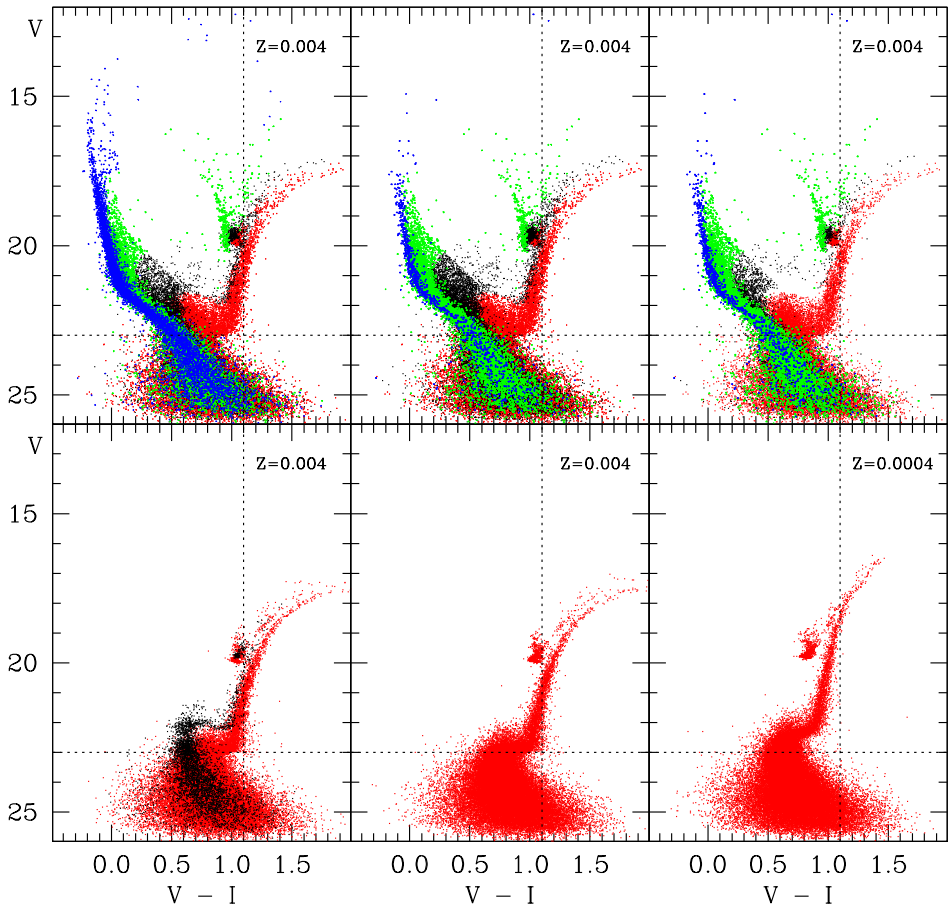}
\caption{The effect of the SFH on the theoretical CMD of a 
hypothetical galactic region 
with (m-M)$_0$=19, E(B-V)=0.08, and with the photometric errors and 
incompleteness typical of HST/WFPC2 photometry. All the shown synthetic
CMDs contain 50000 stars and are based on the Padova models 
(\cite[Fagotto et al. 1994a]{F94a}, \cite[Fagotto et al. 1994b]{F94b}) with the
labelled metallicities.
Top-central panel: the case of a SFR constant from 13 Gyr ago to the present 
epoch.
Top-left panel: the effect of adding a burst 10 times stronger in the last
20 Myr to the constant SFR. The CMD has a much brighter and thicker blue plume.
Top-right panel: same constant SFR as in the
first case, but with a quiescence interval between 3 and 2 Gyrs ago; a gap appears
in the CMD region corresponding to stars 2-3 Gyr old, which are completely
missing. 
Bottom-central panel: SF activity only between 13 and 10 Gyr ago with Z=0.004.
Bottom-right panel: SF activity only between 13 and 10 Gyr ago with Z=0.0004:
notice how colour and luminosity of turnoff, subgiant and red giant branches
differ from the previous case.
Bottom-left panel: SF activity between 13 and 11 Gyr ago, followed by a second
episode of activity between 5 and 4 Gyr ago: a gap separates the two populations
in the CMD, but less evident than in the top-right panel case, when the
quiescent interval was more recent.
}
\label{syn1}       % Give a unique label
\end{figure}

To visualize how the SFH affects the CMD morphology,  a few representative
cases are displayed in Fig.\,\ref{syn1}.  The six panels of the Figure show the
effect of different SFHs on the synthetic CMD of a hypothetical galactic region with 
number of resolved individual stars, photometric errors, blending 
and incompleteness factors typical of a region in the SMC 
imaged with HST/WFPC2. If the SFH of the studied region has been one of the
following six cases, then, according to stellar evolution models, the CMD of 
its  stars is one of
those shown  in Fig.\,\ref{syn1}. The top three
panels show examples of CMDs typical of late-type galaxies, with ongoing or
recent star formation activity. 
If the star formation rate (SFR) has been constant for all the galaxy
lifetime, the CMD of the region is expected to have the morphology of the
top-central panel, with a prominent blue plume mostly populated by main-sequence
(MS) stars and an equally prominent red plume resulting from the overposition
of increasingly bright and massive stars in the red giant branch (RGB), 
asymptotic giant branch (AGB) and red supergiant phases. 
At intermediate colours, for decreasing brightness, stars in the 
blue loops and subgiant phases are visible, as well stars at
the oldest MS turnoff (MSTO) and on the faint MS of low mass stars. Stars of all ages
are present, from those as old as the Hubble time to the brightest ones a few
tens Myr old.

If we leave the SFH unchanged except for the addition of
a burst ten times stronger concentrated in the last 20 Myr, the 
CMD (top-left panel) has a much brighter and more populated blue plume, now
containing also stars a few Myr old.
In the top-right panel the same constant SFR as in the first case is assumed,
but with a quiescent interval between 3 and 2 Gyrs ago: a gap is clearly visible
in the CMD region corresponding to the age of the missing stars. 

The three bottom panels of Fig.\,\ref{syn1}
show CMDs typical of early-type galaxies, whose SF
activity is concentrated at the earliest epochs. If only one SF episode has
occurred from 13 to 10 Gyr ago, with a constant metallicity Z=0.004 as in the
top panel cases, the resulting CMD is shown in the bottom-central panel. If the
SF has occurred at the same epoch, but with a metallicity ten times lower, the
evolutionary phases in the resulting CMD (bottom-right panel) have colours and
luminosities quite different from the previous case. Finally, the bottom-left
panel shows the case of two bursts, the first from 13 to 11 Gyr ago and the
second from 5 to 4 Gyr ago. The gap corresponding to the quiescent interval is
evident in the CMD, although not as much as the more recent gap of the
top-right panel.

The tight dependence of the CMD morphology on the SFH is the cornerstone of the
synthetic CMD method, which consists in comparing the observational CMD of a
galactic region with synthetic CMDs, such as those of Fig.\,\ref{syn1},
 created via Monte~Carlo extractions on
stellar evolution tracks or isochrones for a variety of SFHs, IMFs, 
binary fractions and
age-metallicity relations (see e.g. \cite[Tosi et al. 1991]{T91}, 
\cite[Tolstoy 1996]{T96}, 
\cite[Greggio et al. 1998]{G98}, \cite[Aparicio \& Gallart 2004]{ag04} for 
detailed descriptions of different procedures). 
The synthetic CMDs take into account the number of stars, photometric errors, 
incompleteness and blending factors of the observational 
CMD (or portions of it). Hence, a combination of assumed parameters is acceptable only 
if the resulting synthetic CMD reproduces all
the features of the observational one: morphology, colours, luminosity 
functions, number of stars in specific evolutionary phases.
The method does not provide unique solutions, but significantly reduces 
the possible SFH scenarios. 

At its first applications to photometric data 
from ground-based, moderate size telescopes the synthetic CMD method
demonstrated its power, showing that even in tiny galaxies such as Local Group 
dwarf
irregulars (dIrrs) the SFH varies from one region to the other and that their
star formation regime is rather continuous, with long episodes of moderate
activity, separated by short quiescent intervals (the so-called gasping regime,
 \cite[Ferraro et al. 1989]{F89}, \cite[Tosi et al. 1991]{T91}, 
 \cite[Marconi et  al. 1995]{M95}, \cite[Gallart et al. 1996]{Ga96}, 
 \cite[Tolstoy 1996]{T96}) and not the bursting regime (short episodes of
 strong SF activity separated by long quiescent phases) that most people
 attributed to late-type dwarfs at the time.

When the first non-aberrated images were acquired with HST, the impressive
improvement in the achievable photometric resolution and depth, and the
corresponding quantum leap in the quality of the CMDs, triggered a worldwide
burst of interest in the derivation of the SFHs of nearby galaxies and in the
synthetic CMD method. Many people developped their own procedures and 
to date a large fraction of Local Group galaxies have had the SFH of at least
some of their regions derived with the synthetic CMD method. 
Nowadays, in LG galaxies it is possible to resolve individual stars down to 
faint/old objects in all galactic regions and we can thus infer the SFHs over 
long lookback times $\tau$ (up to the Hubble time), with an average time resolution
around (0.1--0.2)$\tau$.

\begin{figure}[!ht]
\begin{center}
\includegraphics[width = 13cm, height = 6cm]{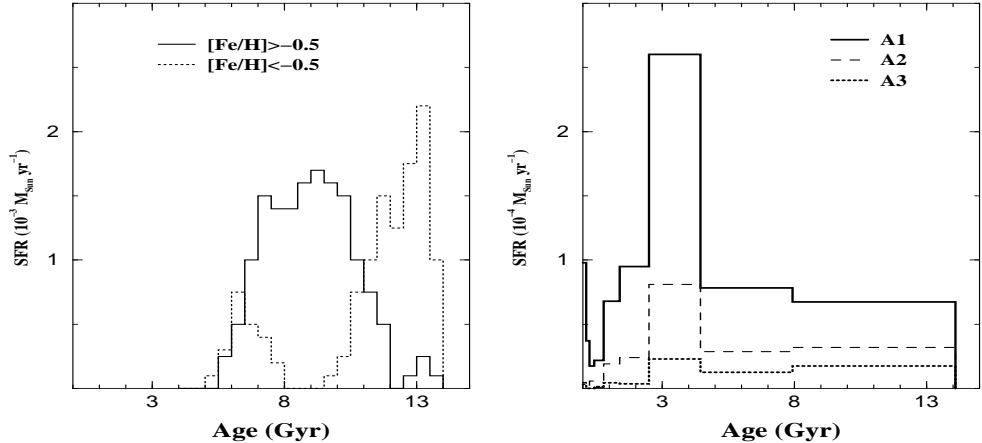}
\end{center}
\caption{SFH from CMDs in (small) regions of the two LG external spirals. 
The left-hand panel shows the SFH of first HST/ACS field of M31 studied by 
\cite{Br06}, who divided the stars according to their metallicity. The
right-hand panel shows the SFHs of the three HST/ACS fields of M33, A1, A2 and 
A3, studied by \cite{Barker07}.}
\label{spiral}
\end{figure}

\section{Star Formation Histories of Local Group Galaxies}

The current census of LG galaxies with SFH derived in some of their regions 
with the synthetic CMD method
 is impressive. SFHs have been inferred from CMDs of
 both the two external spiral galaxies,
M31 and M33, the two Magellanic Clouds, LMC and SMC, 
a dozen dIrrs, 5 transition type dwarfs and about 20 early-type dwarfs
(dwarf spheroidals, dSphs, and dwarf ellipticals, dEs).

In M31, long HST/ACS exposures have allowed \cite{Br08} 
 to resolve stars fainter than the oldest MSTO in
three regions and derive their SFH back to the earliest epochs. They find a 
fairly continuous activity through the whole lifetime of Andromeda. The SFH in
the first M31 field studied by Brown is shown in the left-hand panel of
Fig.\,\ref{spiral}: if the SFH resulting from both metal poor and rich stars is
considered, it turns out to have been rather constant.
In M33, HST/ACS imaging has allowed \cite[Barker et al. (2007)]{B07} to study
three different regions, again resolving their oldest stars. The resulting SFH
(right-hand panel of Fig.\,\ref{spiral}) 
clearly differs from one region to the other and shows significant bumps and
gasps over a rather continuous mode. In all the three regions was the SF
activity already in place a Hubble time ago. It is apparent that the SF activity
in the M31 field has been both stronger and more constant than in M33.

\begin{figure}[!ht]
\begin{center}
\includegraphics[width = 13cm, height = 6cm]{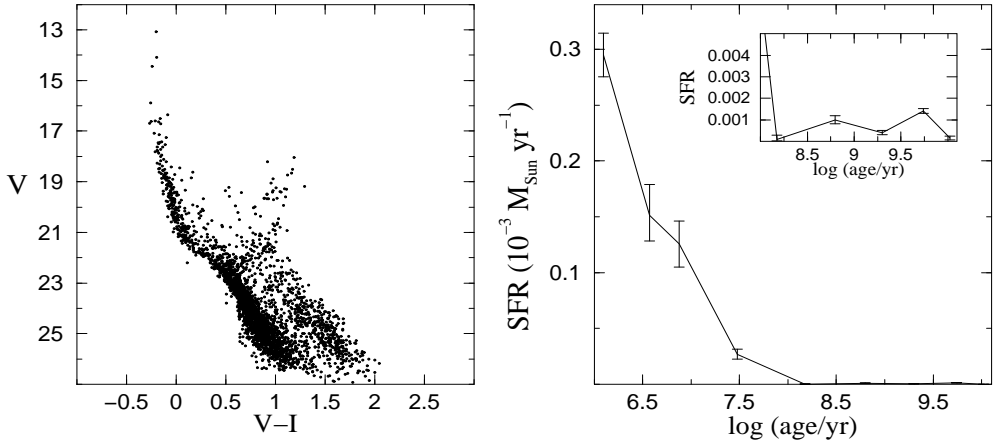}
\end{center}
\caption{ Left-hand panel: CMD of the HST/ACS field around the young 
cluster NGC~602 in the SMC. The red sequence of pre-MS stars 
is  easily recognizable parallel to the lower MS. The bright blue plume contains the young cluster   
stars while the lower MS is only populated by field stars, since the
cluster stars with mass below $\sim$1 $M_{\odot}$ 
haven't yet had time to reach it.
Right-hand panel: corresponding SFH as derived with the
synthetic CMD method (\cite[Cignoni et al. 2009]{Cignoni09}).
 The oldest part of the SFH is zoomed-in in
the upper right inset.}
\label{sfh_602}
\end{figure}

The SFHs of several regions of the Magellanic Clouds have been studied by
a number of authors, both from space and from ground 
(e.g. \cite[Holtzman et al. 1999]{Holtzman99}, 
\cite[Dolphin et al. 2001]{Dolphin01},
\cite[Smecker-Hane et al. 2002]{SH02},
\cite[Harris \& Zaritsky 2004]{HZ04}, 
\cite[Chiosi et al. 2006]{C06}, 
\cite[Noel et al. 2007]{Noel07}, \cite[Cignoni et al. 2009]{Cignoni09}). 
Their proximity makes the
oldest stars visible also from ground, with the advantage of fields of
view larger than those of the HST cameras. \cite{HZ04} even covered the whole SMC.
On the other hand, the exquisite
spatial resolution of HST is necessary to resolve and study the fainter stars in
crowded regions, such as those of the star forming clusters. 
While stars at the oldest MSTOs and subgiant branches are the
unique means to firmly establish the SFH at the earliest epochs,
pre-MS stars are precious tools to study the details of the most
recent SFH in terms of time and space behaviour 
(\cite[Cignoni et al. 2009]{Cignoni09}). 
The SMC regions of intense recent star formation can provide key
information on the star formation 
mechanisms in environments with metallicity much
lower than in any Galactic star forming region. 
Fig.~\ref{sfh_602} shows the CMD of the young cluster NGC~602 in the
Wing of the SMC, observed with HST/ACS.  Both very young stars (either
on the upper MS or still on the pre-MS) and old stars are found. The
SFH of the cluster and the surrounding field is also shown, 
revealing that the cluster has formed most of its
stars around 2.5~Myr ago, while the surrounding field has formed stars
continuously since the earliest epochs.
All the studies on the MC fields have found that the SFHs of their different regions
differ from one another in the details (e.g. epoch of activity peaks, enrichment
history, etc.) but are always characterized by a gasping regime, i.e. a rather
continuous activity since the earliest epochs, but with significant peaks 
and gasps. In the LMC a clear difference has been found between the SFH of field
stars and of star clusters, the latter showing a quiescence phase, several Gyr
long, absent in the field.
 
\begin{figure}[!ht]
\begin{center}
\includegraphics[width = 13cm, height = 6.5cm]{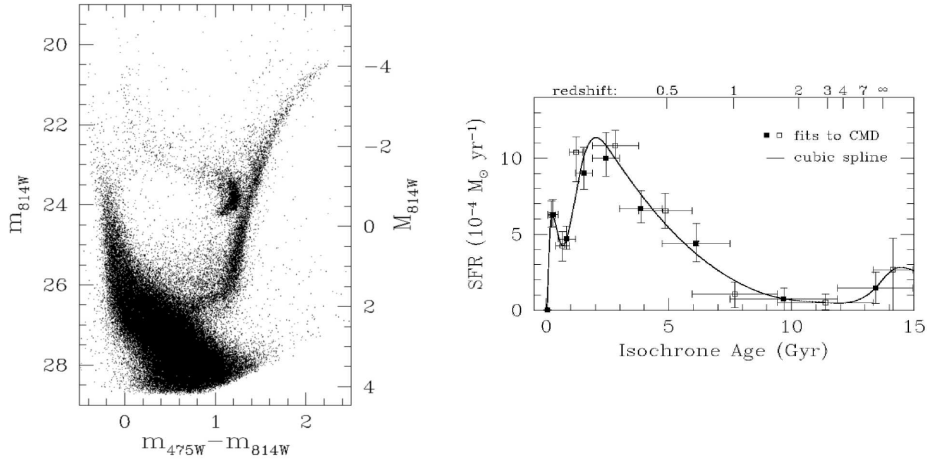}
\end{center}
\caption{CMD and SFH of Leo~A as derived by \cite{Cole07} from HST/ACS data.
Notice the impressive depth and tightness of the CMD, allowing to infer the SFH
even at the earliest epochs.}
\label{leoa}
\end{figure}

Dwarf irregulars were the first systems to which synthetic CMD analyses 
were applied.  HST has had a large impact on studies of these systems. 
The high spatial resolution of its cameras have allowed 
\cite[Dohm-Palmer et al. (1998)]{DP98} and 
\cite[Dohm-Palmer et al. (2002)]{DP02} to 
spatially resolve and measure the SF activity over the last 0.5 Gyr in 
all the sub-regions 
of the dIrrs Gr8 and Sextans A, close to the borders of the LG.  
The resulting space and time distribution of the SF, with lightening and fading of 
adjacent cells, is
intriguingly reminiscent of the predictions of the stochastic self-propagating
SF theory proposed by \cite[Seiden, Schulman, \& Gerola(1979)]{SSG79}
 30 years ago. 
 The HST/ACS is currently providing the deepest and tighter
 CMDs of dIrrs ever obtained, likely to remain unequalled for a very long
 time. These spectacular CMDs reach well below the oldest MSTO
 and allow
 the derivation of the SFH back to a Hubble time ago. The first of such
 impressive studies is that of  Leo A  (\cite[Cole et al. 2007]{Cole07}), 
 whose CMD and 
 SFH are plotted in Fig.\,\ref{leoa}. In Leo~A 
  the star formation activity was present, although quite low, at the earliest 
 epochs, and 90\% of the activity occurred in the last 8 Gyr, with the main peak
 around 2 Gyr ago and a secondary peak a few hundreds Myr ago. 
Once again, the results obtained so far
show that dIrrs experience a rather continuous star formation since the earliest
epochs, but with significant peaks and gasps. 

To find SFHs peaked at earlier epochs one needs to look at early-type dwarfs: dEs,
dSphs and even transition-type dwarfs clearly underwent their major activity
around or beyond 10 Gyr ago. The latter also have significat activity at recent
epochs (e.g. Young et al. 2007).  The former have few (or no) episodes of moderate activity in the last
several Gyrs (e.g. \cite[Smecker-Hane et al. 1996]{SH96}, 
\cite[Hurley-Keller et al. 1998]{Hurley98}, 
\cite[Hernandez, Gilmore \& Valls-Gabaud 2000]{Hernandez00a}, \cite[Dolphin 2002]{Dolphin02},  
\cite[Dolphin et al. 2005]{Dolphin05}).

The beautiful CMDs from Carme Gallart's L-CID HST program on 6 dwarfs of
different type (two dIrrs, two dSphs and two transition type, 
see Hidalgo et al. this volume) promise to provide SFHs of unprecedented
time resolution for external galaxies. 
Another interesting project is trying to treat homogeneously all the LG
galaxies observed with the HST/WFPC2, deriving the CMDs of their resolved
populations in a self-consistent way 
(\cite[Holtzman, Afonso \& Dolphin 2006]{Holtzman06})
and the corresponding SFH with the same technique and assumptions (Dolphin et al.
in preparation, see also \cite[Dolphin et al. 2005]{Dolphin05}). Homogeneous data sets
and analyses are valuable to obtain a uniform overview of the properties of the
different galaxies in the LG.

\begin{figure}[!ht]
\begin{center}
\includegraphics[width = 13cm]{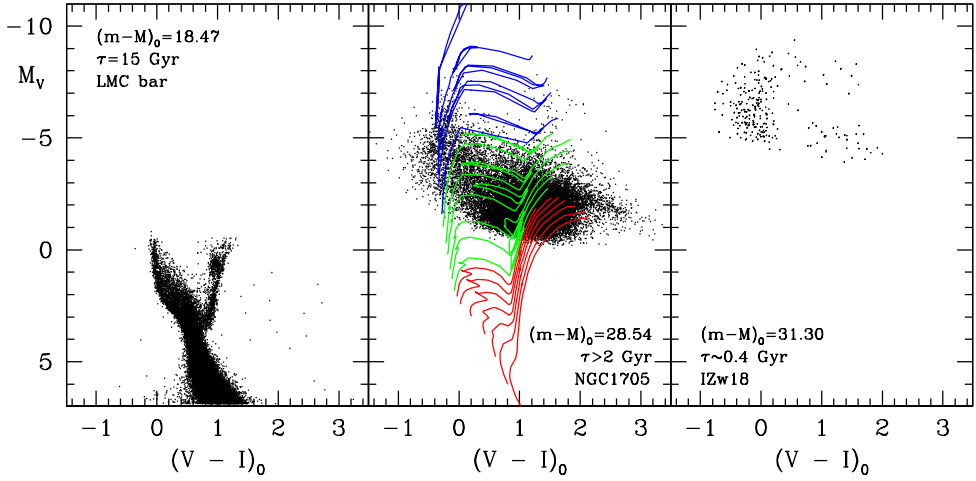}
\end{center}
\caption{Effect of distance on the resolution of individual stars and on the
corresponding lookback time $\tau$ for the SFH. CMD in absolute magnitude and
colour of systems
observed with the HST/WFPC2 and analysed with the same techniques, but
at different distances; from left to right: 50 Kpc (LMC bar), 5.1 Mpc (NGC1705) 
and 18 Mpc 
(IZw18). The central panel also shows stellar evolution tracks from \cite{F94b}
for reference: red lines refer to low-mass stars, green lines to intermediate
mass stars, and blue lines to massive stars.
}
\label{dist}
\end{figure}

\section{Star Formation Histories of galaxies outside the Local Group}
 
In galaxies beyond the LG, distance makes crowding more severe, and even 
HST cannot resolve stars as faint as the  MSTO of old
populations. The higher the distance, the worse the crowding conditions, and the
shorter the lookback time $\tau$ reachable even with the deepest, highest resolution
photometry.  Depending on distance and intrinsic crowding, the reachable 
$\tau$  in galaxies more than 1 Mpc away ranges from several 
Gyrs (in the best cases, when the RGB or even the HB
 are clearly identified), to several hundreds Myr (when AGB  stars are 
 recognized), to a few tens Myr (when only the brightest
supergiants are resolved). 

\begin{figure}[b]
\begin{center}
\includegraphics[width = 6cm]{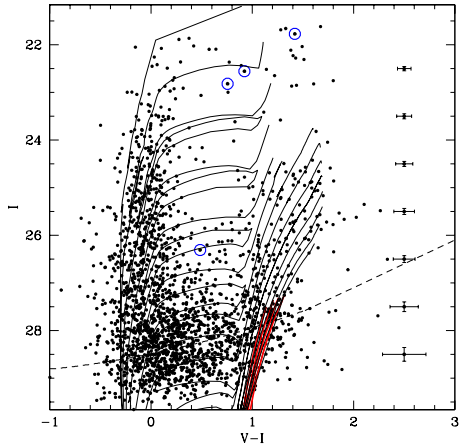}
\end{center}
\caption{CMD of IZw18, obtained from HST/ACS imaging
(\cite[Aloisi et al. 2007]{Aloisi07}). Overimposed are the Z=0.0004 isochrones
by \cite{Bertelli94} with the RGB in red. Also shown is the
average position of the 4 classical Cepheids with reliable light-curves obtained
from these data.
}
\label{IZw18_acs}
\end{figure}

The effect of distance on the possibility of resolving individual stars, and
therefore on the reachable $\tau$, is shown in Fig.\,\ref{dist}, 
where the CMDs
obtained from WFPC2 photometry of three late-type galaxies are shown: the LMC
bar (\cite[Smecker-Hane et al. 2002]{Sme02}), with a distance modulus of 
18.47 (50 kpc) and a CMD reaching several mags
below the old MSTO; NGC1705 (\cite[Tosi et al. 2001)]{To01}, with distance 
modulus 28.54 (5.1 Mpc) and a CMD
reaching a few mags below the tip of the RGB; and IZw18 
(\cite[Aloisi, Tosi \& Greggio 1999]{Al99}), with the new distance
modulus 31.3 (18 Mpc) derived by \cite[Aloisi et al. (2007]{Aloisi07}).
Notice that the latter modulus is inferred  from the periods and luminosities of
a few classical Cepheids measured from HST/ACS data which also allowed us to reach the
RGB, but the WFPC2 data shown in Fig.\,\ref{dist} allow to reach only the AGB. The
CMD obtained from the ACS is shown in Fig.\,\ref{IZw18_acs}.

Since the Local Group doesn't host all types of galaxies, with the notable and
unfortunate absence of both the most and the least evolved ones (ellipticals and
Blue Compact Dwarfs, BCDs, respectively), a few people have tackled the 
challenging task of 
deriving the SFH of more distant galaxies. In spite of the larger uncertainties
and the shorter lookback time, these studies have led to quite interesting
results, which wouldn't have been possible without HST.

\begin{figure}[!ht]
\begin{center}
\includegraphics[width = 13 cm]{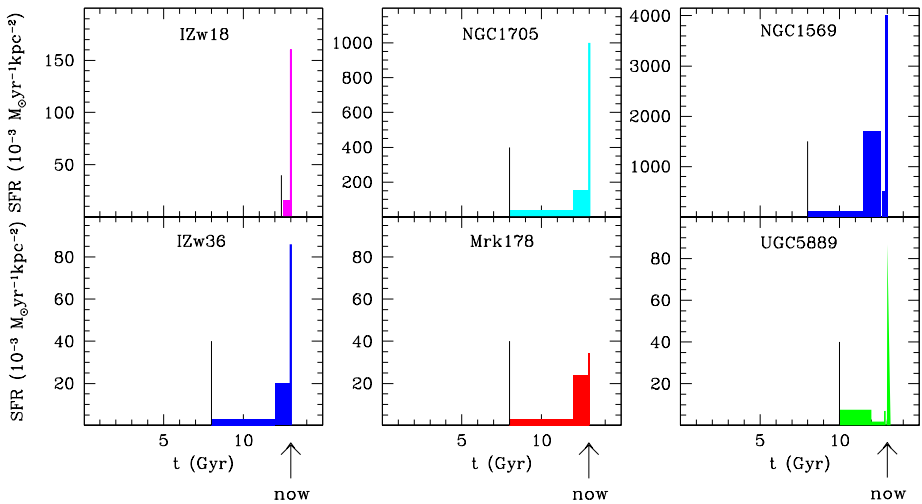}
\end{center}
\caption{SFHs in late-type galaxies derived with the synthetic CMD method. 
In all panels the SF rate per unit area as a function of time is plotted. The
thin vertical line in each panel indicates the reached lookback time. 
References: NGC~1569, \cite{G98,An05};  
NGC~1705, \cite{An03}; IZw18, \cite{Al99}; IZw36, \cite{sl01};
Mrk178, \cite{sl00}; UGC~5889, \cite{Va05}.
}
\label{sfall}
\end{figure}

First of all, all the galaxies, including BCDs, where individual stars have 
been resolved by HST,
 and the SFH has been derived with the synthetic CMD method, have
turned out to be already active at the lookback time reached by the photometry
(see e.g. \cite[Lynds et al. 1998]{Lynds98}, \cite[Aloisi, et al. 1999]{Al99}, 
\cite[Schulte-ladbeck et al. 2000]{sl00}, 
\cite[Schulte-Ladbeck et al. 2001]{sl01}, \cite[Annibali et al. 2003]{An03},
\cite[Rejkuba, Greggio \& Zoccali 2004]{Rejkuba04}, 
\cite[Vallenari Schmidtobreik \& Bomans 2005]{Va05}). 
None of them appears to be experiencing now its first star
formation activity, including the most metal poor ones, such as SBS1415 and
IZw18 (see \cite[Aloisi et al. 2005]{Al05} and \cite[Aloisi et al. 2007]{Al07}). 
Fig.\,\ref{sfall}
sketches the SFHs derived by various authors for some of the starburst dwarfs
studied so far and one low surface brightness dwarf, UGC~5889. 
The lookback time is indicated and in all cases stars with that
age were detected.

All the late-type dwarfs of Fig.\,\ref{sfall}
present a recent SF burst, which is what let people discover them in
spite of the distance, and none of them exhibits long quiescent phases within the
reached $\tau$.  It is interesting to notice that the SFH of the low surface 
brightness dwarf UGC~5889 (\cite[Vallenari et al. 2005]{Va05}) is
also qualitatively similar to that of starburst dwarfs, except that the SFR is
definitely moderate. In all the shown galaxies the strongest SF episodes 
 are overimposed over a rather continuous, moderate SF, already in place at the
 $\tau$ reached by the photometry. Indeed, no one has ever found yet a galaxy 
 without stars as old as $\tau$ from the
 CMDs of the resolved populations. 
 
 The SF rate differs significantly 
from one galaxy to the other. The two most powerful bursts measured so far 
 are the recent ones in NGC~1705 and NGC~1569, with SFR per unit area 
a factor 10-100 higher than in the other starbursting dwarfs studied through
their CMDs. Intriguingly, the strongest of all is not a BCD, but the dwarf
irregular NGC~1569, suggesting that the morphological classification of
these faint small galaxies was possibly affected by their distance and the
capability of resolving their shape with ground-based small telescopes, at the
time of their discovery. Had it
been at 20 Mpc, NGC~1569 would have probably been classified as a BCD.

\section{Discussion}

From the comparison of the SFHs of starburst dwarfs with those of
Local Group dwarf irregulars, one can see that in both cases the SF regime is
rather continuous (gasping), with two main differences: starburst dwarfs always
have the strongest SF episode at recent epochs, while the current SF
activity of local irregulars is not necessarily the highest peak. Leo A, 
with the main SF peak a few Gyr ago (Fig.\,\ref{leoa}) is quite 
typical.  
On the other hand, it is interesting to notice that the SF of the SMC region 
around the very young cluster NGC~602 (Fig.\,\ref{sfh_602}), host of HII regions, 
shows  time
distribution and current rate per unit area similar to those
(Fig.\,\ref{sfall}) of starburst 
dwarfs (once called extragalactic HII regions). The former however involves a
small area, corresponding  to a tiny fraction of the SMC, while the latter are
global behaviours, referring to the whole galaxy.

By comparing with each other the SFHs derived from the CMDs of (few, small) 
regions of the LG spirals (Fig.\,\ref{spiral}), one is tempted to speculate over a
possible dependence of the SFHs on their morphological type and luminosity
class. M31 (SA b I-II) seems to have had very continuous, almost constant SF, 
since the earliest epochs. The solar neighbourhood of the MW (SAB bc II-III)
also shows a rather continuous SF regime, but with larger differences between
the rate of the SF peaks and dips, see e.g. \cite{Rocha97}, \cite{Hernandez00b}
and \cite{Cignoni06}. 
M33 (SA c III) definitely has significant
bumps and gasps over its SFH, with a distribution of SFR with time almost
undistinguishable from that derived for the late-type dwarf Leo A. One could
then argue that the later the morphological type and the lower the luminosity
class of the spirals, the more similar their SFH to those of late-type dwarfs.

Aside from speculations, the general results drawn from all the SFHs derived 
so far for galaxies with CMDs studies can be summarized as follows:

\begin{itemize}
\item Evidence of long interruptions in the SF activity is found only in 
early-type galaxies;

\item Few early-type dwarfs have experienced only one episode of SF activity
concentrated at the earliest epochs: many show instead extended or
recurrent SF activity;

\item  No galaxy currently at its first SF episode has been found yet;

\item  No frequent evidence of strong SF bursts is found in late-type dwarfs;

\item There is no significant difference in the SFH of dIrrs and BCDs, except
for the current SFR.
\end{itemize}

\smallskip\noindent{\bf Acknowledgements}
The Symposium organizers and IAU are gratefully acknowledged for partial
financial support. I thank M. Cignoni and A. Cole for preparing  
figures {\it ad hoc} for this paper, and A. Dolphin and C. Gallart for the 
SFH plots shown in advance of publication.
Some of the results described here have been obtained thanks to frutiful, 
recurrent and pleasant collaborations with A. Aloisi, 
L. Angeretti, F. Annibali, M. Cignoni, L. Greggio, A. Nota, and E. Sabbi. 

%\bibliographystyle{Astronomy} % style aa.bst
%\bibliography{araamtetvhi}

\begin{thebibliography}{}

\bibitem[Aloisi et al. (1999)]{Al99}
{Aloisi, A., Tosi, M., \& Greggio, L.} 1999, \textit{AJ}, 118, 302
\bibitem[Aloisi et al. (2007)]{Al07}
{Aloisi, A., Clementini, G., Tosi, M., Annibali, F., Contreras, R. et al.} 
2007, \textit{ApJ}, 667, L151
\bibitem[Aloisi et al. (2005)]{Al05}
{Aloisi, A., van der Marel, R.P., Mack, J., Leitherer, C., Sirianni, M., \& 
Tosi, M.} 2005, \textit{AJ}, 631, L45
\bibitem[Angeretti et al. (2005)]{An05}
{Angeretti, L., Tosi, M., Greggio, L., Sabbi, E., Aloisi, A.,
	Leitherer, C.}  2005, \textit{AJ}, 129, 2203
\bibitem[Annibali et al. (2003)]{An03}
{Annibali, F., Greggio, L., Tosi, M., Aloisi, A., \& Leitherer, C.} 2003, 
\textit{AJ}, 126, 2752	
\bibitem[Aparicio \& Gallart(2004)]{ag04}
{Aparicio, A., \& Gallart, C.} 2004,  \textit{AJ}, 128, 1465
\bibitem[Barker et al. (2007)]{Barker07}
{Barker, M.K., Sarajedini, A., Geisler, D., Harding, P., Schommer, R.}
2007, \textit{AJ} 133, 1138
\bibitem[Bertelli et al. (1994)]{Bertelli94}
{Bertelli, G., Bressan, S., Chiosi, C., Nasi, E.} 1994, \textit{A\&AS}, 106, 275
\bibitem[Bertelli \& Nasi (2001)]{Bertelli01}
{Bertelli, G. \& Nasi, E.} 2001, \textit{AJ}, 121, 101
\bibitem[Brown (2006)]{Br06}
{Brown, T.M.} 2006, in \textit{The Local Group as Astrophysical Laboratory},
M.Livio, T.M.Brown eds, STScI Symp.Ser. 17 (CUP), p.111
\bibitem[Brown et al.(2008)]{Br08}
{Brown, T.M.,  Beaton, R., Chiba, M., Ferguson, H.C., Gilbert, K.M. et al.}
 2008, \textit{ApJ},658, L121
\bibitem[Chiosi et al. (2006)]{C06}
{Chiosi, E., Vallenari, A., Held, E.V., Rizzi, L., Moretti, A.} 2006,
\textit{A\&A}, 452, 179
\bibitem[Cignoni et al. (2006)]{Cignoni06}
{Cignoni, M., Degl'Innocenti, S., Prada Moroni, P.G., Shore, S.N.} 2006,
\textit{A\&A}, 459, 783
\bibitem[Cignoni et al. (2009)]{Cignoni09}
{Cignoni, M., Sabbi, E., Nota, A., Tosi, M., Degl'Innocenti, S.,  Prada
Moroni, P., Angeretti, L., Carlson, L., Gallagher, J., Meixner, M., Sirianni, 
M., Smith, L.J.} 2009, \textit{AJ}, in press
\bibitem[Cole et al. (2007)]{Cole07}
{Cole, A.A., Skillman, E.D., Tolstoy, E., Gallagher, J.S., Aparicio, A., et al.}
2007, \textit{ApJ}, 659, L17
\bibitem[Dohm-Palmer et al.(1998)]{DP98}
{Dohm-Palmer, R.C., Skillman, E.D., Gallagher, J.S., Tolstoy, E., Mateo, M.,
Dufour, R.J., Saha, A., Hoessel, J., Chiosi, C.} 1998, \textit{AJ}, 116, 1227
\bibitem[Dohm-Palmer et al.(2002)]{DP02}
{Dohm-Palmer, R.C., Skillman, E.D., Mateo, M., Saha, A., Dolphin, A., Tolstoy,
E., Gallagher, J.S., Cole, A.A.} 2002, \textit{AJ}, 123, 813
\bibitem[Dolphin (2002)]{Dolphin02}
{Dolphin, A.E.} 2002, \textit{MNRAS}, 332, 91
\bibitem[Dolphin (2005)]{Dolphin05}
{Dolphin, A.E., Weisz, D.R., Skillman, E.D., Holtzman, J.A.} 2005,
\textit{astro-ph/0506430}
\bibitem[Dolphin et al. (2001)]{Dolphin01}
{Dolphin, A.E., Walker, A.R., Hodge, P.W., Mateo, M., Olszewski, W.W., Schommer,
R.A., Suntzeff, N.B.} 2001, \textit{ApJ}, 562, 303 	
\bibitem[Fagotto et al. (1994a)]{F94a}
{Fagotto, F., Bressan, A., Bertelli, G., Chiosi, C.} 1994a, \textit{A\&AS}, 104,
365
\bibitem[Fagotto et al. (1994b)]{F94b}
{Fagotto, F., Bressan, A., Bertelli, G., Chiosi, C.} 1994b, \textit{A\&AS}, 105,
29
\bibitem[Ferraro et al.(1989)]{F89}
{Ferraro, F. R., Fusi Pecci, F., Tosi, M.,  Buonanno, R.} 1989, \textit{MNRAS}, 241, 433
\bibitem[Gallart et al.(1996)]{Ga96}
{Gallart, C., Aparicio, A., Bertelli, G., Chiosi, C.} 1996, \textit{AJ}, 112, 1950
\bibitem[Greggio et al.(1998)]{G98}
{Greggio, L., Tosi, M., Clampin, M., De Marchi, G., Leitherer, C.,
 Nota, A., Sirianni, M.} 1998, \textit{ApJ}, 504, 725
\bibitem[Harris \& Zaritsky (2004)]{HZ04}
{Harris, J. \& Zaritsky, D.} 2004, \textit{ApJ}, 127, 1531
\bibitem[Hernandez, Gilmore \& Valls-Gabaud (2000)]{Hernandez00a}
{Hernandez, X., Gilmore, G., Valls-Gabaud, D.} 2000, \textit{MNRAS}, 317, 831
\bibitem[Hernandez, Valls-Gabaud \& Gilmore  (2000)]{Hernandez00b}
{Hernandez, X., Valls-Gabaud, D., Gilmore, G.} 2000, \textit{MNRAS}, 316, 605
\bibitem[Holtzman et al. (1999)]{Holtzman99}
{Holtzman, J.A., Gallagher, J.S., Cole, A.A., Mould, J.R., et al.} 1999, 
\textit{AJ}, 118, 2262
\bibitem[Holtzman et al. (2006)]{Holtzman06}
{Holtzman, J.A., Afonso, C., Dolphin, A.E.} 2006, \textit{ApJS}, 166, 534
\bibitem[Hurley-Keller et al. (1998)]{Hurley98}
{Hurley-Keller, D., Mateo, M., Nemec, J.} 1998, \textit{AJ}, 115, 1840
\bibitem[Lynds et al. (1998)]{Lynds98}
{Lynds, R., Tolstoy, E.,  O'Neil., E.J.Jr., Hunter, D.A.} 1998, \textit{AJ}, 116, 146
\bibitem[Marconi et al.(1995)]{M95}
{Marconi, G., Tosi, M., Greggio, L., Focardi, P.} 1995, \textit{AJ}, 109, 173
\bibitem[Noel et al. (2007)]{N07}
{Noel, N.E.D., Gallart, C., Costa, E., Mendez,
R.A.} 2007, \textit{AJ}, 133, 2037
\bibitem[Rejkuba et al. (2004)]{Rejkuba04}
{Rejkuba, M., Greggio, L., Zoccali, M.} 2004, \textit{A\&AS}, 415, 915
\bibitem[Rocha-Pinto \& Maciel (1997)]{Rocha97}
{Rocha-Pinto, H.J. \& Maciel, W.J.} 1997, \textit{MNRAS}, 289, 882		
\bibitem[Seiden, Schulman, \& Gerola(1979)]{SSG79}
{Seiden, P.E., Schulman, L.S., Gerola, H.} 1979, \textit{ApJ}, 232, 709
\bibitem[Schulte-Ladbeck et al. (2000)]{sl00}
{Schulte-Ladbeck, R.E., Hopp, U., Greggio, L., Crone, M.M.} 2000, 
\textit{AJ}, 120, 1713
\bibitem[Schulte-Ladbeck et al. (2001)]{sl01}
{Schulte-Ladbeck, R.E., Hopp, U.,  Greggio, L., Crone, M.M.,
 Drozdovsky, I.O.} 2001, \textit{AJ}, 121, 3007
\bibitem[Smecker-Hane et al.(2002)]{SH02}
{Smecker-Hane, T.A., Cole, A.A.,  Gallagher, J.S.III;, Stetson, P.B.} 2002, 
\textit{ApJ}, 566, 239
\bibitem[Smecker-Hane et al.(1996)]{SH96}
{Smecker-Hane, T.A., Stetson, P.B., Hesser, J.E., Vandenberg, D.A.} 1996,
in \textit{From stars to galaxies}, \textit{PASP Conf.Ser.}, 98, 328
\bibitem[Tolstoy(1996)]{T96}
{Tolstoy, E.}  1996, \textit{ApJ}, 462, 684
\bibitem[Tosi et al. (1991)]{To91}
{Tosi, M., Greggio, L., Marconi, G., Focardi, P.} 1991, \textit{AJ}, 102, 951
\bibitem[Tosi et al.(2001)]{To01}
{Tosi, M., Sabbi, E., Bellazzini, M., Aloisi, A.,
 Greggio, L., Leitherer, C., Montegriffo, P.} 2001, \textit{AJ}, 122, 127
\bibitem[Vallenari et al.(2005)]{Va05}
{Vallenari,  A., Schmidtobreick, L., \& Bomans, D.J.} 2005, \textit{A\&Ap}, 435, 821
\bibitem[Young et al. (2007)]{Young07}
{Young, L.M., Skillman, E.D., Weisz, D.R., Dolphin, A.E.} 2007, \textit{ApJ},
659, 331
\end{thebibliography}
%\end{document}

\end{document}